# Correlated Hofstadter Spectrum and Flavor Phase Diagram in Magic Angle Graphene

Jiachen Yu[1,2], Benjamin A. Foutty[2,3], Zhaoyu Han[3], Mark E. Barber[1,2], Yoni Schattner[3,4], Kenji Watanabe[5], Takashi Taniguchi[6], Philip Phillips[7], Zhi-Xun Shen[1,2,3,4], Steven A. Kivelson[2,3], Benjamin E. Feldman[2,3,4†]

[1]*Department of Applied Physics, Stanford University, Stanford, CA 94305, USA*
[2]*Geballe Laboratory of Advanced Materials, Stanford, CA 94305, USA*
[3]*Department of Physics, Stanford University, Stanford, CA 94305, USA*
[4]*Stanford Institute for Materials and Energy Sciences, SLAC National Accelerator Laboratory, Menlo Park, CA 94025, USA*
[5]*Research Center for Functional Materials, National Institute for Material Science, 1-1 Namiki, Tsukuba 305-0044, Japan*
[6]*International Center for Materials Nanoarchitectonics, National Institute for Material Science, 1-1 Namiki, Tsukuba 305-0044, Japan*
[7]*Department of Physics and Institute for Condensed Matter Theory, University of Illinois at Urbana-Champaign, Urbana, IL 61801, USA*

[†]email: bef@stanford.edu

**In magic angle twisted bilayer graphene (MATBG), the moiré superlattice potential gives rise to narrow electronic bands[1] which support a multitude of many-body quantum phases[2–12]. Further richness arises in the presence of a perpendicular magnetic field, where the interplay between moiré and magnetic length scales leads to fractal Hofstadter subbands[13]. In this strongly correlated Hofstadter platform, multiple experiments have identified gapped topological and correlated states[14–20], but little is known about the phase transitions between them in the intervening compressible regimes. Here, using a scanning single-electron transistor microscope to measure local electronic compressibility, we simultaneously unveil novel sequences of broken-symmetry Chern insulators (CIs) and resolve sharp phase transitions between competing states with different topological quantum numbers and spin/valley flavor occupations. Our measurements provide a complete experimental mapping of the energy spectrum and thermodynamic phase diagram of interacting Hofstadter subbands in MATBG. In addition, we observe full lifting of the degeneracy of the zeroth Landau levels (zLLs) together with level crossings, indicating moiré valley splitting. We propose a unified flavor polarization mechanism to understand the intricate interplay of topology, interactions, and symmetry breaking as a function of density and applied magnetic field in this system.**

When two monolayer graphene sheets are twisted near a magic angle of 1.1°, spatially modulated interlayer tunneling generates a pair of flat electronic bands at low energies[1]. The combination of quenched kinetic energies and multiple quantum degrees of freedom renders the system prone to various intriguing correlation-driven phases characterized by spontaneously broken symmetry[2–12,21–34]. At zero magnetic field, a cascade of Stoner-type transitions was reported[9,10] to result in a series of changes in flavor polarization and a sawtooth pattern of inverse compressibility as a function of moiré filling factor $\nu$. The same mechanism underlies the recent observation[11,12] of a Pomeranchuk effect at $\nu = \pm 1$.

In a perpendicular magnetic field $B$, moiré systems develop a fractal energy spectrum in the non-interacting limit known as the Hofstadter butterfly[13] when the flux density per moiré unit cell approaches $\Phi/\Phi_0 \sim O(1)$. Whereas the Hofstadter spectrum has only been theoretically calculated at a single-particle level[35–39], the flat electronic bands of MATBG open up new opportunities to investigate Hofstadter butterfly physics in the strongly correlated regime. Recent studies have identified a series of robust correlated CIs stabilized at moderate fields, which have been interpreted in terms of flavor-polarized states in which only subsets of the underlying Chern bands[16,18,19] or Hofstadter subbands[14,15,17] are occupied, with a driving mechanism analogous to quantum Hall (QH) ferromagnetism that is a natural extension of the cascade model. Most studies have focused only on the flavor polarization/degeneracy of the CIs themselves, with limited insight between the incompressible states. Moreover, no theoretical or experimental work has determined the full strongly correlated Hofstadter spectrum, in MATBG or any other solid-state system.

Here we measure the local inverse electronic compressibility $d\mu/dn$ of MATBG using a scanning single-electron transistor (SET). Our measurements reveal an unexpected sequence of broken-symmetry CIs at high fields, demonstrate how they evolve from the sawtooth pattern of electronic compressibility at low fields, and enable us to quantitatively extract the strongly correlated Hofstadter spectrum. We further observe multiple sharp features that exhibit negative $d\mu/dn$, which correspond to flavor phase transitions, some of which are first order and exhibit hysteresis. Because our thermodynamic probe is sensitive in both compressible and incompressible regimes, we are able to pinpoint exactly where phase transitions occur and construct a phase diagram of the number of partially occupied ('active') flavors as a function of $\nu$ and $B$. Finally, we observe robust broken-symmetry QH states emanating from the charge neutrality point (CNP) at all integer Landau level (LL) fillings $\nu_C$ and apparent LL crossings at $\nu_C = \pm 2$. These behaviors suggest moiré valley polarization induced by mirror symmetry ($M_y$) breaking. The nanoscale resolution of the SET renders it less susceptible to spatial inhomogeneity and allows us to disentangle which of these phenomena are robust, and which are sensitive to local details such as twist angle.

**Correlated Broken-Symmetry Hofstadter Butterfly**

An optical micrograph of the sample and schematic of the experimental setup are respectively shown in Fig. 1a,b (see Methods and Supplementary Information). At $B = 0$, $d\mu/dn$ exhibits a sawtooth-like pattern (Fig. 1c), and measurements as a function of position demonstrate similar behavior over a range of twist angles (Extended Data Fig. 1). All integer fillings including the CNP are gapless, indicating that the hBN is not aligned. The data presented below contrast with and complement a recent local compressibility study of an aligned device[34]. In this paper, we focus on the behavior of MATBG in a perpendicular field, where we observe multiple interpenetrating sequences of incompressible states (Fig. 1d). Each can be described by the Diophantine equation[40], $\nu = t(\Phi/\Phi_0) + s$, where $t$ is the sum of the Chern numbers ($C$) of all filled bands, and $s$ the band filling index. Figure 1e shows the experimentally determined Wannier diagram[40].

The incompressible states can be grouped by their intercept at $B = 0$. When $s = 0$, they correspond to LLs originating from the CNP (in this manuscript, we refer to all $s = 0$ LLs as zLLs). We observe all the integer zLL states with $|\nu_C| < 4$, which have comparable strengths and emerge at similar $B$. This requires full symmetry breaking, including lifting the two-fold degeneracy that would arise from the presence of two degenerate moiré valleys, which we discuss in more detail at the end of the manuscript. We also observe a series of LLs emanating from the superlattice (SL) gaps at $|s| = 4$ that point towards lower density. They are much weaker than the zLLs, indicating that the narrow bands are less dispersive close to the SL gaps compared to near the CNP. Finally, gapped states with $0 < |s| < 4$ are CIs. Consistent with previous reports, we find a singly degenerate $s = 3$ sequence, two-fold degenerate $|s| = 2$ sequences, and no CIs emanating from $|s| = 1$ in the low field regime[4,5,14,41], and the strongest CIs satisfy the relation $|s| + |t| = 4$ at high fields[14–20]. These CIs are robust over a finite range of twist angles (Extended Data Fig. 2). They are independently confirmed by the changes in local conductivity probed by scanning microwave impedance microscopy (Extended Data Fig. 3).

Novel correlated CIs and other distinctive features appear at high fields. Nearly all reports on hBN-unaligned samples show CIs emanating away from the CNP with increasing $B$. In contrast, here we observe a robust positive sloped (1, -3) state that forms above $B \sim 6$ T. Similarly, a vertical incompressible (0, -2) state appears above $B \sim 9$ T. Neither of these states have been reported before, and they are striking because they represent reentrant behavior in $s$ as a function of $\nu$ and $B$. In addition, the $s = -1$ CIs follow an unusual hierarchy $t = \{-3, -2, -5, -7\}$ (in descending strength), whereas its electron counterpart has a sequence of $t = \{3, 2, 4, 5\}$. Finally, we observe a CI that extrapolates to a non-integer $s = -1/2$, indicative of translational symmetry breaking[14]. The phenomenology described above is largely reproduced in other independent regions of the sample (Extended Data Fig. 4) and is still robust at $T = 1.7$ K (Extended Data Fig. 5).

The sequences of CIs can be understood in the framework of correlation-driven flavor-polarized Hofstadter subbands[14,15], and they further clarify the subband structure itself. We propose a phenomenological Hofstadter subband diagram of MATBG for a single flavor at finite $B$, schematically illustrated in Fig 2a. The Chern number associated with each subband is constrained by the topology of the underlying bands together with experimentally observed zLL/CI sequences (see Supplementary Information). Near zero energy, doublets of subbands each carry $C = 1$, reflecting the presence of two Dirac nodes in moiré Brillouin zone (mBZ) of the same helicity that are energetically distinguished by the aforementioned moiré valley splitting[29,42,43]. The subbands that extend from the SL band edge carry $C = 1$. Finally, the remaining subbands whose single-particle gaps are small compared to disorder and/or thermal broadening form one large subband that contains the van Hove singularity (vHs). This large subband carries $C = -4$ in order to satisfy the total $C = 0$ constraint of the whole spectrum.

Figure 2b illustrates how the broken symmetry CIs can be constructed from selective filling of the four flavors. The strongest CIs correspond to equal filling of an integer number of flavors up to the largest $C = 1/-1$ gaps[14,15,17] (top panel of Fig. 2b). Other broken symmetry CIs can be derived from these 'parent' CIs by further populating finer subbands in the same active flavors. The bottom panels of Fig. 2b show how this leads to the experimentally observed CI sequence emanating from $s = -1$. Adding charge to a single moiré valley subband of one flavor produces a (-2, -1) state, and removing charge from the highest occupied doublet in one or two flavors leads to the (-5, -1) and (-7, -1) CIs. A more comprehensive explanation of all the CIs we observe is detailed in the Supplementary Information.

Our data indicate a delicate competition between different CIs in the presence of single-particle and interaction-driven symmetry breaking. This is exemplified by the striking feature in Fig. 1e near $\nu = -8/3$, $B = 8.7$ T and the corresponding region in Fig. 1d, where different CIs with $s$ = -4, -3, -2, and -1 all intersect. The behavior near this parameter range also shows strong spatial dependence (Extended Data Fig. 6). We therefore conclude that details such as small twist angle and strain variations can tip the balance between closely competing CIs. The presence or absence of specific broken symmetry CIs involves details of the interaction, various single-particle effects and disorder, and is therefore difficult to unambiguously predict. We note that although multiple combinations of subbands can often produce the right ($t$, $s$), those that require additional numbers of active flavors are likely energetically unfavorable, because they generate both large single particle and interaction energy penalties.

A key benefit of our measurement technique is that we can extract thermodynamic gaps and bandwidths by integrating $d\mu/dn$ to obtain $\mu(n, B)$. This enables us to experimentally determine the interacting Hofstadter spectrum (Fig. 2c) based on the data in Fig. 1d (Extended Data Fig. 9 shows an analogous spectrum from another location). Each dot in the plotted spectrum represents one state of energy $E$ at a given $B$. Hence, the density of dots represents the density of states (DOS), similar to a numerically calculated Hofstadter spectrum. The overall shape of the spectrum generally agrees with the continuum model based single-particle calculations[35-37], which is characterized by the saturation of the $\nu_C = \pm 4$ zLL gaps and a reduction in total bandwidth with increasing $B$. While individual Hofstadter subbands are evident, their details differ from a simple replication of the single-particle spectrum due to the strong correlations present in MATBG. Most notably, we observe additional spectral gaps associated with broken-symmetry CIs at high magnetic fields as well as LL-like CIs emerging from $\nu = \pm 2$ at moderate fields. The spectrum also exhibits a recurring pattern of dense points which reflect vHs-like features where the flavor phase transitions occur (Fig. 2c, black arrows).

The data in Fig. 2c represents the first quantitative measurement of an interacting Hofstadter spectrum. The resulting spectrum is distinct from typical tunneling spectroscopy measurements of DOS in which a tunnel bias is swept while maintaining fixed density. This is because our measurement always probes at the Fermi energy and is therefore sensitive to many-body effects at all plotted energies, whereas electrons generally tunnel above or below the Fermi energy in tunneling spectroscopies. Our approach also avoids tip-induced doping and soft Coulomb gaps that can be present in tunneling experiments.

**First Order Polarization Transitions And Flavor Phase Diagram**

Our measurements also provide detailed information about the phase diagram of flavor occupancy as a function of $\nu$ and $B$. We first discuss the abrupt appearance of the CIs at $(\pm 3, \pm 1)$. Theory predicts that these phase transitions can be first order[33], and transport experiments have indeed shown abrupt onset[14] and hysteresis from other Chern states emanating from $\nu = 1$ (Ref. [5]). In our experiment, they form abruptly in the middle of zLL fan and are flanked by sharply negative $d\mu/dn$ (Fig 3b-c, e-f). Their energy gaps saturate soon after they form (Fig. 3h), whereas the gaps of all other CIs exhibit a more gradual increase near their onset (Fig. 3g; see Supplementary Information for details of thermodynamic gap extraction). Strikingly, we observe hysteresis in $d\mu/dn$ near the critical $B$ and $\nu$ of the $(\pm 3, \pm 1)$ states, which provides the first direct thermodynamic evidence of a first order phase transition between unpolarized/partially-polarized zLLs and broken symmetry CIs. Hysteresis is apparent both when $\nu$ is modified at fixed $B$ (Fig. 3a-f) as well as when $B$ is adjusted at fixed $\nu$ (Extended Data Fig. 7). Interestingly, the hysteretic region coincides with the intersection point of zLLs with $(\pm 3, \pm 1)$, and is not ubiquitous in the entire negative $d\mu/dn$ regions (Fig. 3a,d). We do not observe hysteresis near these transitions at 1.7 K, and no hysteresis was evident in any other parts of the phase diagram even at base temperature. However, we cannot preclude that other phase transitions are also first order but are masked by disorder and/or thermal broadening.

We observe multiple additional features where $d\mu/dn < 0$, many of which occur between incompressible states. Negative $d\mu/dn$ can be a hallmark of interaction-driven ground state phase transitions, where the free energies of different electronic states cross[44,45]. In MATBG, these crossings correspond to flavor polarization transitions, and based on the locations of the negative $d\mu/dn$ features, we can construct a phase diagram that characterizes the number of active flavors in the entire $\nu - B$ plane (Fig. 3i; see Supplementary Information). This phase diagram, which is a central result of this paper, reveals an intricate pattern of phase boundaries, including reentrant behavior in the number of active flavors. The detailed slopes and curvatures of the negative compressibility features are likely related to the underlying spin/orbital moments of the adjacent phases. While a full explanation of their form is beyond the scope of this work, our results inform and constrain microscopic theories that address specific spin and valley ordering throughout the phase diagram.

Our data also clarify the evolution between the cascade of phase transitions at $B = 0$, whose characteristic sawtooth persists to finite fields, and the high-field regime of flavor polarized Hofstadter subbands. An estimated boundary of the sawtooth is superimposed on top of the flavor phase diagram in Fig. 3i (dashed lines). The zLLs extend far beyond $\nu = \pm 1$, through its sawtooth, and only terminate close to $\nu = \pm 2$. The low field CIs with $|s| = 2, 3$ also coexist with the sawtooth over a range of densities. The continuation of the sawtooth in nonzero $B$ and the coexisting zLLs/CIs can be simultaneously understood by considering Stoner ferromagnetism in the Hofstadter subbands near the CNP. This gives rise to a hierarchy of transitions: 'subleading' transitions within active flavors generate zLLs or low field CIs. Upon increasing density toward the vHs, these transitions give way to 'master' transitions which completely reset the number of active flavors and terminates the CIs. These 'master' transitions can be adiabatically continued the cascade of transitions at $B = 0$. Interestingly, this mechanism was recently reported in compressibility measurement on rhombohedral trilayer graphene[45], hinting at the universality of the polarizing mechanism in correlated itinerant electronic systems with multiple internal degrees of freedom.

**Zeroth Landau Levels, CI Degeneracies and Moiré Valley Polarization**

The sequence and strengths of the observed zLLs provide further insight into the symmetry breaking present in our device. Figure 4a shows the zLLs plotted as a function of their filling factor $\nu_C$ and $B$. We resolve all zLLs within $-4 < \nu_C < 4$; surprisingly, both even and odd integers emerge at similar magnetic fields and show similar gap sizes (Fig. 4b-c). This does not show obvious angle dependence over a large spatial region that encompasses a variety of twist angles (Extended Data Figs. 2 and 4), and it contrasts with past work where states at odd integers were absent[3,4,15,46], which has been explained in terms of a two-fold moiré valley degeneracy. Although singly degenerate zLLs have been previously reported[14,16–20,41], no consensus has been reached on the driving mechanism. A second important experimental finding is the dips in the energy gaps at $\nu_C = \pm 2$ that occur near $B = 4$ T (Fig. 4c). This behavior indicates zLL crossings and it is also reproduced in multiple locations in the sample (Extended Data Figs. 4 and 10).

Multiple theoretical mechanisms have been proposed that could potentially lift the zLL degeneracy, and we consider each in turn below. The density dependence of compressibility and magnitudes of the energy gaps at $\nu_C = \pm 4$ are consistent with a Dirac dispersion near the CNP, rendering the scenario where band-touching occurs at the center of mBZ unlikely[17]. Two types of symmetry breaking can lift the zLL degeneracy[29,38]: $C_{3z}$ and $M_y$. Breaking $C_{3z}$ alone can halve the zLL degeneracy but is predicted to involve vHs physics and occur at finite $B$. Moreover, no level crossings are expected in this scenario. We therefore conclude that $M_y$ breaking is the most likely explanation for the observed zLL behavior[17].

In MATBG, $M_y$ can be broken by external effects such as heterostrain[29,42,43], and can also be caused or further enhanced by interactions[47]. This would energetically distinguish the two Dirac nodes[29,42,43] in each mBZ by an amount $\Delta_\nu$ (Extended Data Fig. 8a). Two quartets of zLLs are expected to emanate from each respective Dirac cone with correlation gaps that scale linearly with applied field due to inter-flavor repulsion $U\nu_i\nu_j$ between flavors $i$ and $j$[33] (see Supplementary Information). Extended Data Fig. 8b illustrates the corresponding evolution of the zLL free energy, which results in crossings[17] between states associated with different moiré valleys. This naturally explains the closing and reopening of the $\nu_C = \pm 2$ gaps as a function of $B$, and also matches the observation that the odd $\nu_C$ gaps have similar magnitudes at high fields. Within this model, the effective $U$ can be extracted from the $B$ dependence of the zLL gaps (Figs. 4b,c) to be ~10 meV (see Supplementary Information), which is consistent with estimates from earlier thermodynamic measurements[11,15]. Other crossings are predicted but are not apparent in our data; their absence can be explained by differences in inter-flavor splitting (Extended Data Fig. 8d,e; see Supplementary Information). We note that this mechanism does not require the two Dirac nodes to reside at the mBZ corners, and is therefore still valid in the presence of $C_{3z}$ breaking. Furthermore, such moiré valley splitting also provides a natural explanation for the singly degenerate sequence of CIs emanating from $\nu = 3$, which cannot otherwise be reconciled with the presence of two degenerate moiré valleys contributing to the Hofstadter spectrum[48].

**Conclusion**

In conclusion, we have experimentally determined the interacting Hofstadter spectrum and phase diagram of flavor occupancy in MATBG, and have shown that a unifying flavor symmetry breaking mechanism can capture the physics at multiple energy and field scales. A microscopic understanding of the specific spin and valley ordering in various regions of the phase diagram remains an exciting open question to be explored, and the close competition between ground states that we observe suggest that external tuning, e.g. by strain, can be used as control knobs to switch between different correlated and topological ground states. More generally, we anticipate that the methods demonstrated here can be applied to map the phase diagrams and probe Hofstadter physics across the broad and rapidly growing set of strongly correlated moiré systems[45,49].

**References**

1. Bistritzer, R. & MacDonald, A. H. Moire bands in twisted double-layer graphene. *Proc. Natl. Acad. Sci.* **108**, 12233–12237 (2011).

2. Cao, Y. *et al.* Correlated insulator behavior at half-filling in magic-angle graphene superlattices. *Nature* **556**, 80–84 (2018).

3. Cao, Y. *et al.* Unconventional superconductivity in magic-angle graphene superlattices. *Nature*

**556**, 43–50 (2018).

4. Yankowitz, M. *et al.* Tuning superconductivity in twisted bilayer graphene. *Science* **363**, 1059–1064 (2019).

5. Lu, X. *et al.* Superconductors, orbital magnets and correlated states in magic-angle bilayer graphene. *Nature* **574**, 653–657 (2019).

6. Sharpe, A. L. *et al.* Emergent ferromagnetism near three-quarters filling in twisted bilayer graphene. *Science* **365**, 605–608 (2019).

7. Serlin, M. *et al.* Intrinsic quantized anomalous Hall effect in a moiré heterostructure. *Science* **367**, 900–903 (2020).

8. Cao, Y. *et al.* Nematicity and competing orders in superconducting magic-angle graphene. *Science* **372**, 264–271 (2021).

9. Zondiner, U. *et al.* Cascade of phase transitions and Dirac revivals in magic-angle graphene. *Nature* **582**, 203–208 (2020).

10. Wong, D. *et al.* Cascade of electronic transitions in magic-angle twisted bilayer graphene. *Nature* **582**, 198–202 (2020).

11. Saito, Y. *et al.* Isospin Pomeranchuk effect in twisted bilayer graphene. *Nature* **592**, 220–224 (2021).

12. Rozen, A. *et al.* Entropic evidence for a Pomeranchuk effect in magic-angle graphene. *Nature* **592**, 214–219 (2021).

13. Hofstadter, D. R. Energy levels and wave functions of Bloch electrons in rational and irrational magnetic fields. *Phys. Rev. B* **14**, 2239–2249 (1976).

14. Saito, Y. *et al.* Hofstadter subband ferromagnetism and symmetry-broken Chern insulators in twisted bilayer graphene. *Nat. Phys.* **17**, 478–481 (2021).

15. Park, J. M., Cao, Y., Watanabe, K., Taniguchi, T. & Jarillo-Herrero, P. Flavour Hund's coupling, Chern gaps and charge diffusivity in moiré graphene. *Nature* **592**, 43–48 (2021).

16. Nuckolls, K. P. *et al.* Strongly correlated Chern insulators in magic-angle twisted bilayer

graphene. *Nature* **588**, 610–615 (2020).

17. Choi, Y. *et al.* Correlation-driven topological phases in magic-angle twisted bilayer graphene. *Nature* **589**, 536–541 (2021).

18. Das, I. *et al.* Symmetry-broken Chern insulators and Rashba-like Landau-level crossings in magic-angle bilayer graphene. *Nat. Phys.* (2021) doi:10.1038/s41567-021-01186-3.

19. Wu, S., Zhang, Z., Watanabe, K., Taniguchi, T. & Andrei, E. Y. Chern insulators, van Hove singularities and topological flat bands in magic-angle twisted bilayer graphene. *Nat. Mater.* **20**, 488–494 (2021).

20. Tomarken, S. L. *et al.* Electronic Compressibility of Magic-Angle Graphene Superlattices. *Phys. Rev. Lett.* **123**, 046601 (2019).

21. Xie, Y. *et al.* Spectroscopic signatures of many-body correlations in magic-angle twisted bilayer graphene. *Nature* **572**, 101–105 (2019).

22. Kerelsky, A. *et al.* Maximized electron interactions at the magic angle in twisted bilayer graphene. *Nature* **572**, 95–100 (2019).

23. Jiang, Y. *et al.* Charge order and broken rotational symmetry in magic-angle twisted bilayer graphene. *Nature* **573**, 91–95 (2019).

24. Choi, Y. *et al.* Electronic correlations in twisted bilayer graphene near the magic angle. *Nat. Phys.* **15**, 1174–1180 (2019).

25. Saito, Y., Ge, J., Watanabe, K., Taniguchi, T. & Young, A. F. Independent superconductors and correlated insulators in twisted bilayer graphene. *Nat. Phys.* **16**, 926–930 (2020).

26. Stepanov, P. *et al.* Untying the insulating and superconducting orders in magic-angle graphene. *Nature* **583**, 375–378 (2020).

27. Liu, X. *et al.* Tuning electron correlation in magic-angle twisted bilayer graphene using Coulomb screening. *Science* **371**, 1261–1265 (2021).

28. Stepanov, P. *et al.* Competing zero-field Chern insulators in Superconducting Twisted Bilayer Graphene. *ArXiv201215126 Cond-Mat* (2020).

29. Po, H. C., Zou, L., Vishwanath, A. & Senthil, T. Origin of Mott Insulating Behavior and Superconductivity in Twisted Bilayer Graphene. *Phys. Rev. X* **8**, 031089 (2018).

30. Bultinck, N. *et al.* Ground State and Hidden Symmetry of Magic-Angle Graphene at Even Integer Filling. *Phys. Rev. X* **10**, 031034 (2020).

31. Kang, J. & Vafek, O. Strong Coupling Phases of Partially Filled Twisted Bilayer Graphene Narrow Bands. *Phys. Rev. Lett.* **122**, 246401 (2019).

32. Liu, J., Liu, J. & Dai, X. Pseudo Landau level representation of twisted bilayer graphene: Band topology and implications on the correlated insulating phase. *Phys. Rev. B* **99**, 155415 (2019).

33. Lian, B. *et al.* Twisted bilayer graphene. IV. Exact insulator ground states and phase diagram. *Phys. Rev. B* **103**, 205414 (2021).

34. Pierce, A. T. *et al.* Unconventional sequence of correlated Chern insulators in magic-angle twisted bilayer graphene. *ArXiv210104123 Cond-Mat* (2021).

35. Bistritzer, R. & MacDonald, A. H. Moiré butterflies in twisted bilayer graphene. *Phys. Rev. B* **84**, 035440 (2011).

36. Moon, P. & Koshino, M. Energy spectrum and quantum Hall effect in twisted bilayer graphene. *Phys. Rev. B* **85**, 195458 (2012).

37. Hejazi, K., Liu, C. & Balents, L. Landau levels in twisted bilayer graphene and semiclassical orbits. *Phys. Rev. B* **100**, 035115 (2019).

38. Zhang, Y.-H., Po, H. C. & Senthil, T. Landau level degeneracy in twisted bilayer graphene: Role of symmetry breaking. *Phys. Rev. B* **100**, 125104 (2019).

39. Lian, B., Xie, F. & Bernevig, B. A. Landau level of fragile topology. *Phys. Rev. B* **102**, 041402 (2020).

40. Wannier, G. H. A Result Not Dependent on Rationality for Bloch Electrons in a Magnetic Field. *Phys. Status Solidi B* **88**, 757–765 (1978).

41. Uri, A. *et al.* Mapping the twist-angle disorder and Landau levels in magic-angle graphene. *Nature* **581**, 47–52 (2020).

42. Bi, Z., Yuan, N. F. Q. & Fu, L. Designing flat bands by strain. *Phys. Rev. B* **100**, 035448 (2019).

43. Parker, D. E., Soejima, T., Hauschild, J., Zaletel, M. P. & Bultinck, N. Strain-induced quantum phase transitions in magic angle graphene. *ArXiv201209885 Cond-Mat* (2020).

44. Eisenstein, J. P., Pfeiffer, L. N. & West, K. W. Negative compressibility of interacting two-dimensional electron and quasiparticle gases. *Phys. Rev. Lett.* **68**, 674–677 (1992).

45. Zhou, H. *et al.* Half and quarter metals in rhombohedral trilayer graphene. *ArXiv 2104.00653 Cond-Mat* (2021).

46. Arora, H. S. *et al.* Superconductivity in metallic twisted bilayer graphene stabilized by WSe2. *Nature* **583**, 379–384 (2020).

47. Huang, K. S., Han, Z., Kivelson, S. A. & Yao, H. Pair-Density-Wave in the Strong Coupling Limit of the Holstein-Hubbard model. *ArXiv 2103.04984 Cond-Mat* (2021).

48. Kang, J., Bernevig, B. A. & Vafek, O. Cascades between light and heavy fermions in the normal state of magic angle twisted bilayer graphene. *ArXiv 2104.01145 Cond-Mat* (2021).

49. Kennes, D. M. *et al.* Moiré heterostructures as a condensed-matter quantum simulator. *Nat. Phys.* **17**, 155–163 (2021).

50. Wang, L. *et al.* One-Dimensional Electrical Contact to a Two-Dimensional Material. *Science* **342**, 614–617 (2013).

51. Chen, G. *et al.* Evidence of a gate-tunable Mott insulator in a trilayer graphene moiré superlattice. *Nat. Phys.* **15**, 237–241 (2019).

## Methods

**Device fabrication**

The MATBG stack was fabricated using standard dry transfer technique[50]. A monolayer graphene flake was pre-cut by a conductive AFM probe[25,26,51] in contact mode, with an ac excitation of 10 V$_{p-p}$ at 50 kHz, in order to minimize tearing-induced strain. The two isolated flakes were rotated 1.15° relative to each other before stacking and encapsulated by top (27 nm) and bottom (40 nm) hBN. Few-layer graphite was used as the bottom gate. Metal electrodes were subsequently patterned using standard e-beam lithography techniques to form edge contacts[50].

**Scanning SET measurements**

The SET sensor was fabricated by evaporating aluminum on the apex of a pulled quartz rod. The size of the apex, and thus the lateral dimension of the SET, is estimated to be 50 - 80 nm. It was brought to < 50 nm above the MATBG sample surface, resulting in an overall spatial resolution of about 100 nm. The scanning SET measurements were performed in a Unisoku USM 1300 SPM with a customized microscope head. An ac excitation $V_g = $ 4 - 8 mV at $f_g = $ 911.999 Hz was applied to the back gate, and an ac excitation $V_{2d} = $ 1 mV at $f_{2d} = $ 773.777 Hz was applied to the sample. We then measure inverse compressibility $d\mu/dn \propto I_g/I_{2d}$, where $I_g, I_{2d}$ are demodulated from the SET current through the SET probe using standard lock-in techniques. A dc offset voltage $V_{2d}$ is further applied to the sample to maintain maximum sensitivity of the SET and minimize tip-induced doping (see Supplementary Information). All data presented are taken at 330 mK unless otherwise specified.

**Gate capacitance and twist angle determination**

The sample gate capacitance (conversion between density and applied gate voltage) was calibrated by measuring the slope of zLLs, which yielded a value consistent with geometric considerations. The twist angle is then given by the relation $\theta(r) = a\sqrt{\sqrt{3}n_s(r)/8}$ where $a = 0.246$ nm is graphene's lattice constant, and $n_s(r)$ is the carrier density at full filling.

**Background subtraction**

A finite geometric capacitance between the tip and backgate electrode can contribute to an overall constant spurious background in measured $d\mu/dn$. To estimate this parasitic signal, we compare the measured $d\mu/dn$ of the higher dispersive bands at $|\nu| > 4$ in the contact area and at the center of the device, where the twist angle is slightly larger (~ 1.14 - 1.2°). The latter has a lower background due to better screening of the back gate and larger separation from its electrode. We found a small but finite difference of $8 \times 10^{-12}$ meV cm², which we subtract from all the presented data. Differences in $d\mu/dn$ from differences in the details of the dispersive bands is small, and therefore will not affect our observations. The background contribution is still likely to be finite even at the center of the sample given the small size of the device, so we conclude that we are likely under-subtracting the true parasitic background signal. This will enlarge the measured bandwidth by a small ($< 5\%$) amount but has a negligible effect on the extracted gap sizes.

**Scanning Microwave Impedance Microscopy (MIM) Measurement**

MIM measurements were performed in a ³He cryostat with a 12 T superconducting magnet and custom scanner incorporating Attocube nanopositioners. The MIM probe, an etched tungsten wire, was attached to a quartz tuning fork for topographic sensing and scans were taken with the tip held approximately 20 nm above the sample's surface. MIM measures changes in admittance between the tip and sample at GHz frequencies which can be related to changes in local conductivity and permittivity in the sample. The measurements reported here were carried out at 6.8 GHz. At GHz frequencies the tip and sample are strongly capacitively coupled, enabling sub-surface sensing without requiring additional electrical contacts on the sample. MIM operates in the near-field limit and the spatial resolution is dictated by the tip diameter, ~ 100 nm, rather than the microwave wavelength.

**Data availability**

The data that supports the findings of this study are available from the corresponding authors upon reasonable request.

**Code availability**

The codes that support the findings of this study are available from the corresponding authors upon reasonable request.

**Acknowledgements**

We acknowledge helpful discussions with D. Goldhaber-Gordon and O. Vafek. This work was supported by the QSQM, an Energy Frontier Research Center funded by the U.S. Department of Energy (DOE), Office of Science, Basic Energy Sciences (BES), under Award # DE-SC0021238. B.E.F. acknowledges a Stanford University Terman Fellowship and an Alfred P. Sloan Foundation Fellowship. K.W. and T.T. acknowledge support from the Elemental Strategy Initiative conducted by the MEXT, Japan, Grant Number JPMXP0112101001, and JSPS KAKENHI Grant Number JP20H00354. S.A.K. acknowledges support from the Department of Energy, Office of Basic Energy Sciences, under Contract No. DEAC02-76SF00515. B.A.F. acknowledges a Stanford Graduate Fellowship. M.E.B acknowledges support from the Marvin Chodorow postdoctoral fellowship of the Applied Physics department at Stanford University. Y.S. was supported by the Gordon and Betty Moore Foundation's EPiQS Initiative through grants GBMF 4302 and GBMF 8686. P.W.P. acknowledges partial support from NSF grant DMR-2111379. Part of this work was performed at the Stanford Nano Shared Facilities (SNSF), supported by the National Science Foundation under award ECCS-1542152.

**Author Contribution**

J.Y, B.A.F. and B.E.F. designed and conducted the scanning SET experiments. M.E.B. and Z.-X.S. designed and conducted the MIM experiments. B.A.F. fabricated the sample. Z.H., Y.S., S.A.K., and P.W.P. contributed to theoretical analysis. K.W. and T.T. provided hBN crystals. All authors participated in discussions and in writing of the manuscript.

**Competing Interests**

Z.-X.S. is a co-founder of PrimeNano Inc., which licensed the Microwave Impedance Microscopy (MIM) technology from Stanford University for commercial instruments.

**Figure 1**

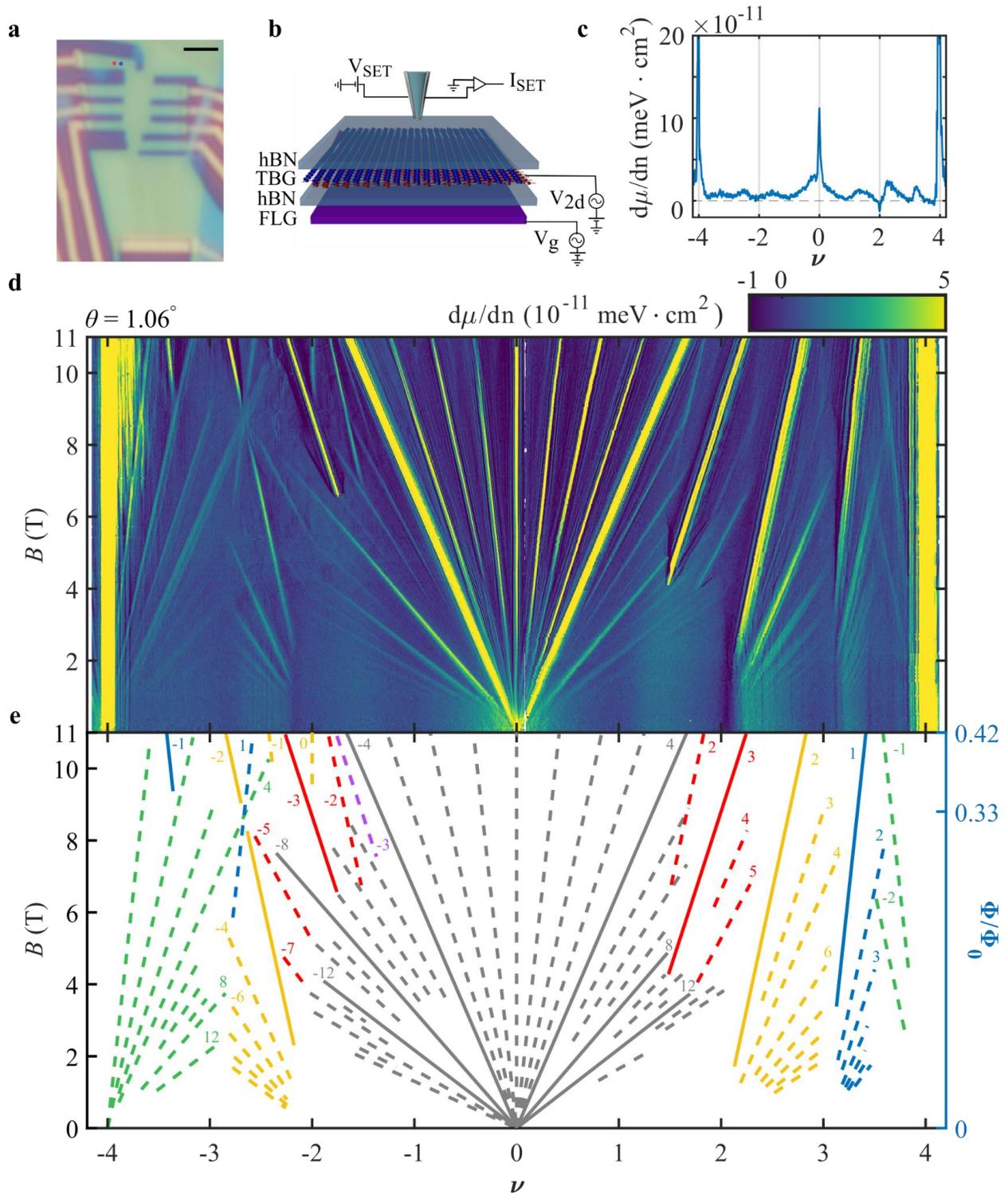

**Fig. 1 | Device geometry and local electronic compressibility of MATBG. a,** Optical micrograph of the MATBG device. Scale bar, 2 $\mu$m. **b,** Schematic of the measurement setup (see Methods). **c,** Inverse

electronic compressibility $d\mu/dn$ measured as a function of moiré filling factor $\nu$ at perpendicular magnetic field $B = 0$. **d**, $d\mu/dn$ as a function of $\nu$ and $B$, measured at the location marked by the red dot in **a**. Data is truncated at $-1 \times 10^{-11} \sim 5 \times 10^{-11}$ meV cm² in the colormap. **e,** Incompressible (gapped) states identified from **d**. Selected states are labeled by their slopes $t$. Gray, red, yellow, blue, green and purple colors correspond to states with different integer and fractional intercepts (-1/2), respectively. Solid lines represent the strongest state in each group. For the zLLs they denote multiples of $|t| = 4$.

**Figure 2**

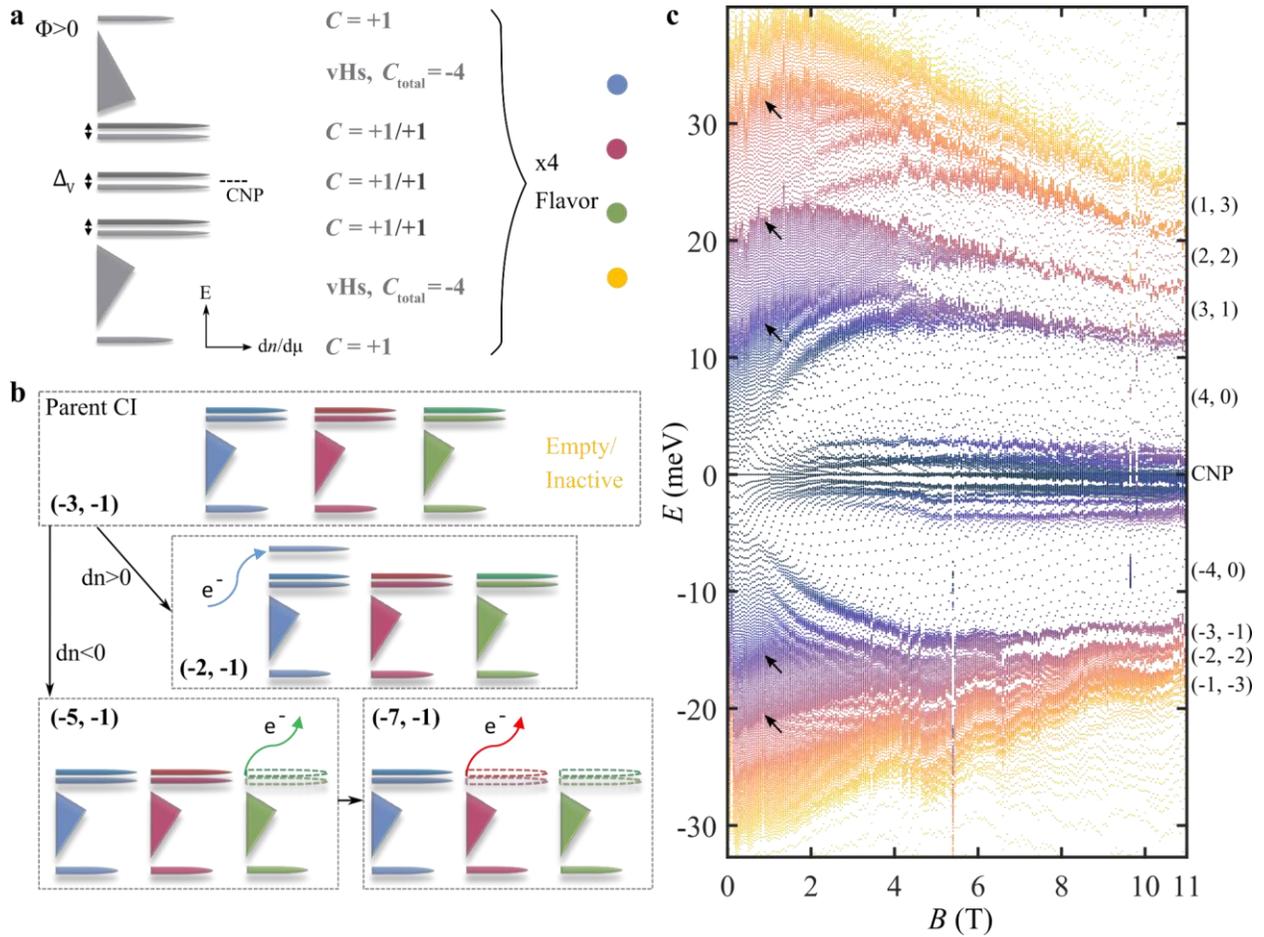

**Fig. 2 | Flavor polarization and correlated Hofstadter Spectrum. a,** Schematic of a single-flavor Hofstadter subband diagram at flux per moiré unit cell $\Phi > 0$. The Chern number ($C$) of each subband is labeled on the right; vHs denotes subbands close to the van Hove singularity that are unresolved due to broadening/disorder. The possibility of moiré valley splitting $\Delta_v$ in the central subbands is indicated by doublets with $C = 1$. **b,** Illustration of flavor occupancy of the broken symmetry CI sequence emanating from $\nu = -1$. The upper panel shows the parent CI (-3, -1). The lower panels demonstrate how the experimentally observed sequence can be derived by selectively populating (middle) or emptying (bottom) subbands in a given flavor. **c,** Experimentally determined many-body Hofstadter spectrum. Each dot corresponds to an eigenstate of the Hofstadter spectrum. The dot density thus reflects the many-body density of states (DOS). The colors denote eigenstates (at fixed density $n$) and show how their energies change upon flux insertion. The spectral gaps that correspond to parent CIs and $\nu_C = \pm 4$ zLLs are labeled. Features characteristic of the phase transition cascades are marked by black arrows.

# Figure 3

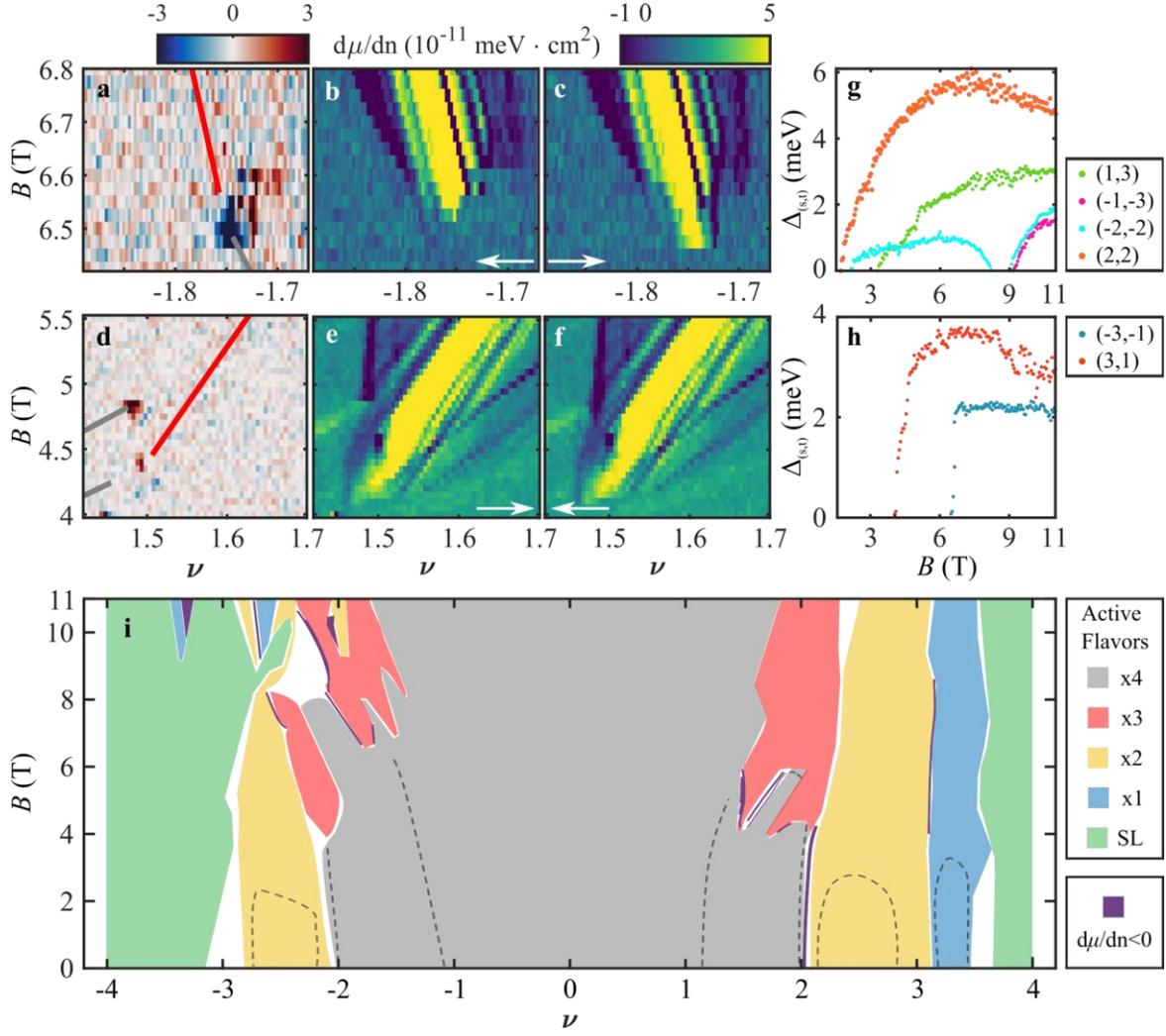

**Fig. 3 | First order phase transitions and flavor phase diagram. a-c,** Hysteresis $d\mu/dn_{inc} - d\mu/dn_{dec}$(**a**) derived from increasing (**b**) and decreasing (**c**) density sweeps near the onset of the (-3, -1) CI. Red and gray lines in (**a**) show the positions of (-3, -1) and (-7, 0) states, respectively. **d-f,** Hysteresis $d\mu/dn_{inc} - d\mu/dn_{dec}$(**d**) derived from increasing (**e**) and decreasing (**f**) density sweeps near the (3, 1) onset transition. Red and gray lines in (**d**) show the positions of (3, 1), (8, 0), and (9, 0) states, respectively. **g-h,** Thermodynamic energy gaps of the predominant $|s| + |t| = 4$ sequence parent CIs as a function of $B$. **i,** Flavor phase diagram of MATBG as a function of $\nu$ and $B$. Gray, red, yellow, blue, denote 4, 3, 2 and 1 active flavors, and green labels 4 active flavors near the superlattice gap. Purple lines and areas denote regions where we observe negative $d\mu/dn$, indicating where phase transitions occur. Dashed lines indicate boundaries of the low-field sawtooth in $d\mu/dn$.

**Figure 4**

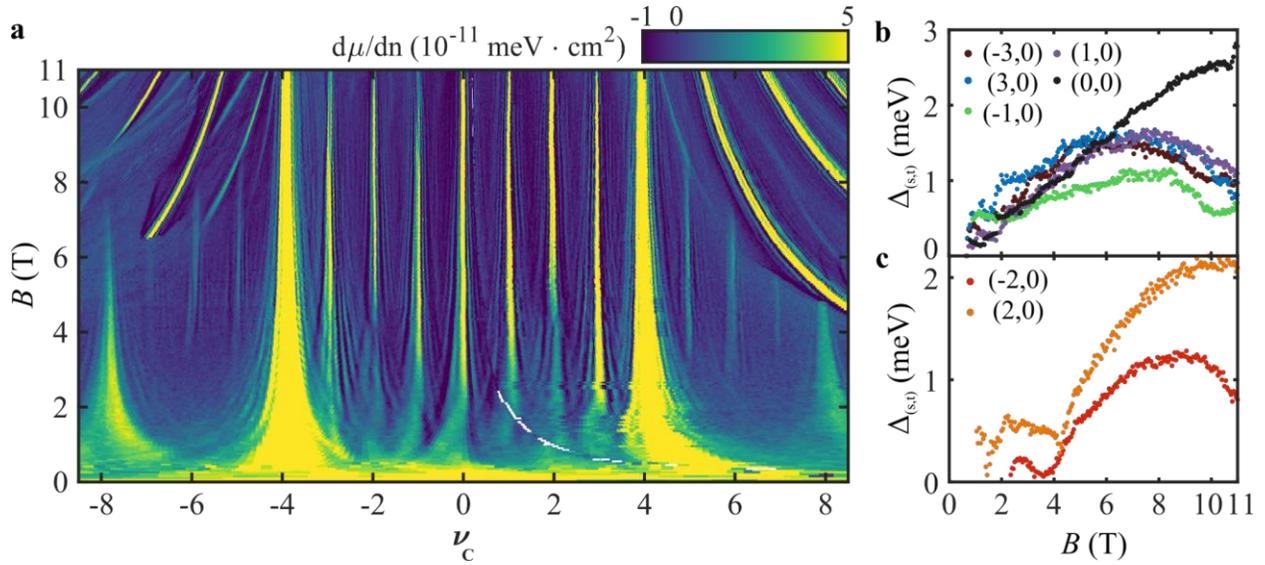

**Fig. 4 | Zeroth Landau levels and moiré valley polarization. a,** Measured $d\mu/dn$ in Fig. 1d replotted as a function of the zLL filling factor $\nu_C$ and $B$. **b-c,** Thermodynamic energy gaps of states at $\nu_C = 0, \pm1, \pm3$ (**b**) and $\nu_C = \pm2$ (**c**). The dips in the gaps of $\nu_C = \pm2$ gaps near 4 T indicate Landau level crossings.

# Extended Data Figure 1

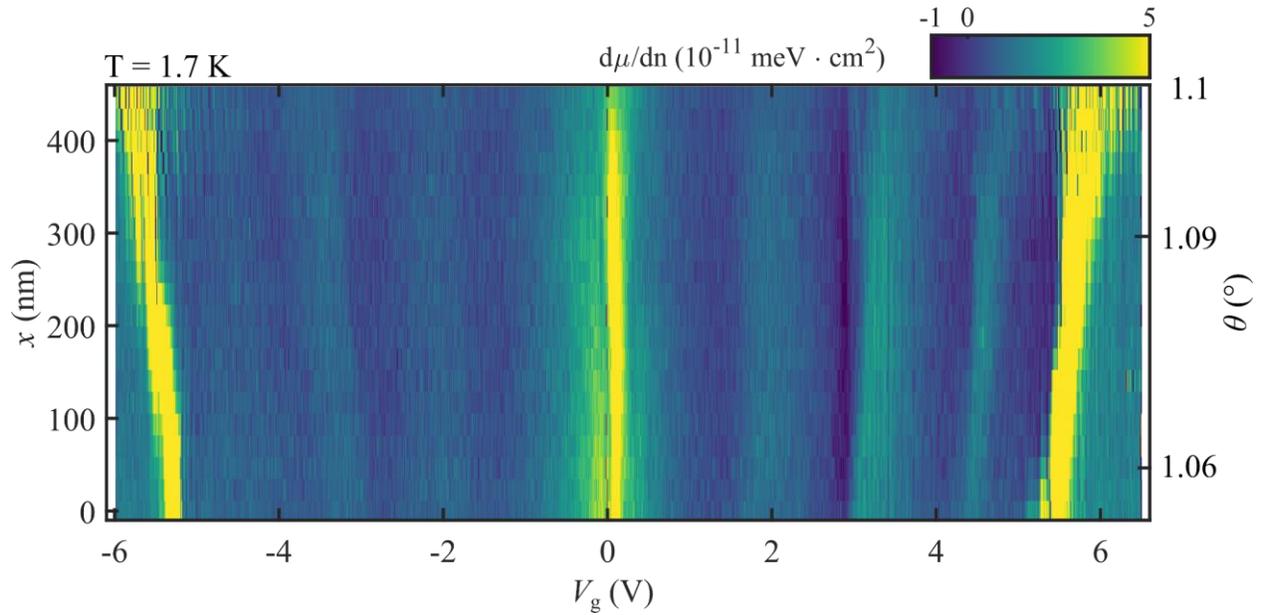

**Extended Data Fig. 1 | Spatial dependence of electronic compressibility at zero field.** Linecut of $d\mu/dn$ as a function of gate voltage ($V_g$) measured along a line in the top left contact area of the device shown in Fig 1a. The Landau fans shown in Fig. 1 and Extended Data Fig. 4 are taken at $x$ = 40 nm and $x$ = 300 nm, respectively. The sawtooth pattern on the electron side generically exhibits a larger amplitude and negative $d\mu/dn$ at the $\nu = 2$ transition.

# Extended Data Figure 2

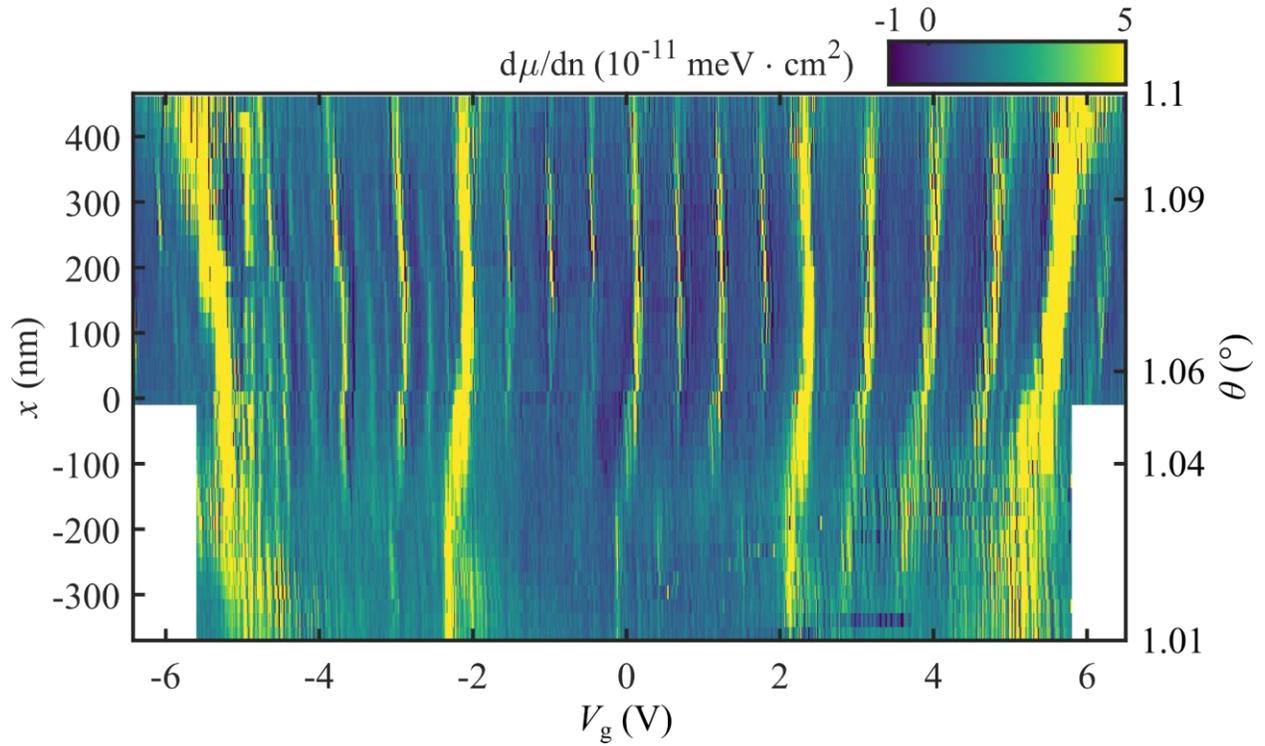

**Extended Data Fig. 2 | Spatial dependence of zLL and CIs.** Linecut of $d\mu/dn$ at 11 T. The positions with $x > 0$ correspond to those shown in Extended Data Fig. 1. While the overall strengths of the broken-symmetry zLLs is spatially dependent, those at odd integers have comparable strengths to those at even integers independent of position.

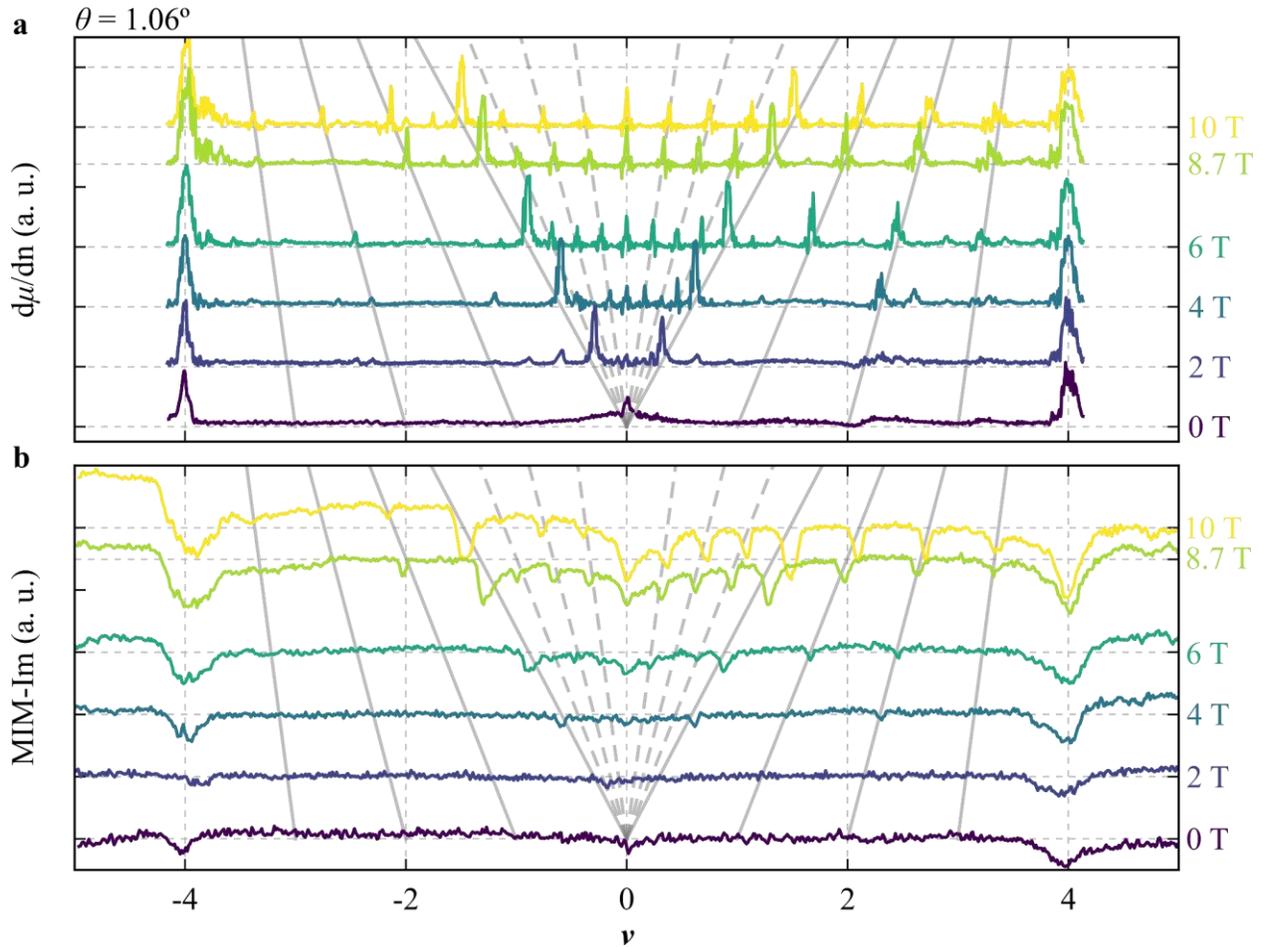

**Extended Data Fig. 3 | Comparison of inverse electronic compressibility and MIM measurements.
a,** Selected $d\mu/dn$ measurements as a function of moiré filling factor $\nu$ at different perpendicular magnetic fields reproduced from Fig 1d. The curves are vertically spaced according to the applied magnetic field and the grey lines indicate the field dependence of the incompressible states identified in Fig 1e. **b,** MIM measurements of the local conductivity at the same sample position as a function of $\nu$ and at the same magnetic fields. A decrease in MIM-Im corresponds to a decrease in conductivity confirming the main CI and zLLs are associated with local resistive behavior. The MIM measurements were taken at $T = 450$ mK.

# Extended Data Figure 4

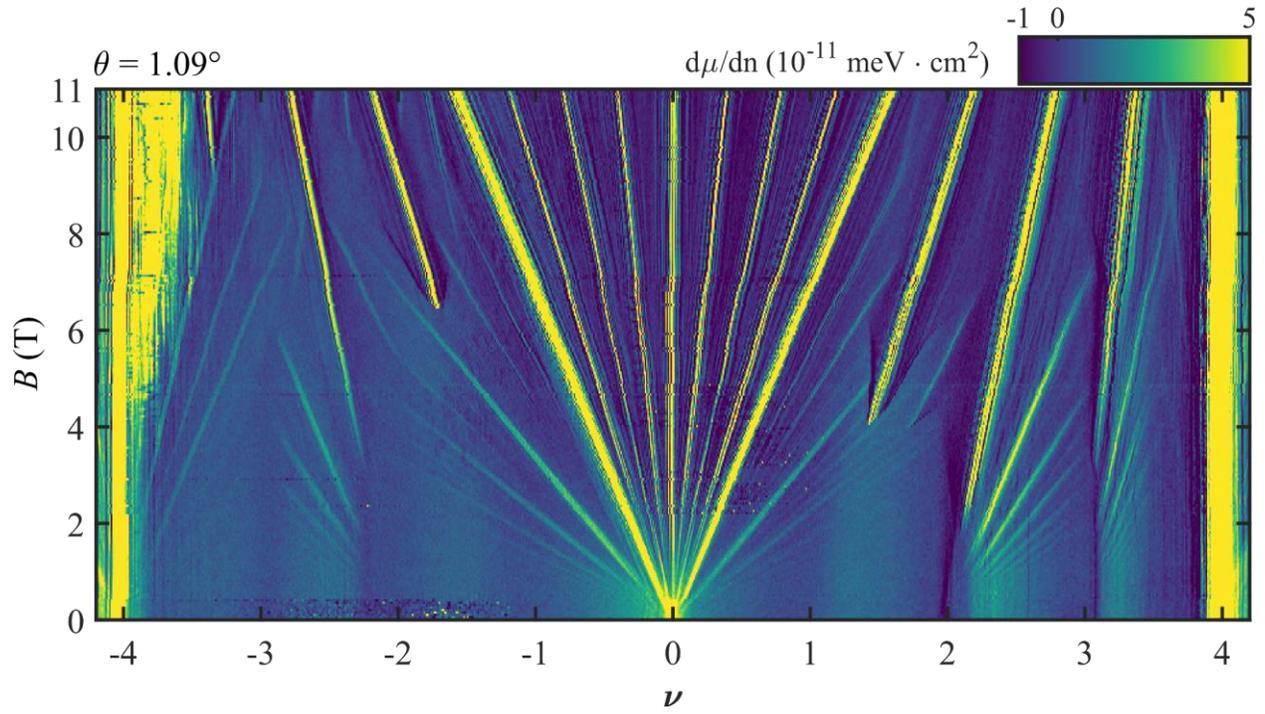

**Extended Data Fig. 4 | Landau fan of CIs at a second location.** $d\mu/dn$ measured at a second location (blue dot in Fig. 1a) with twist angle $\theta = 1.09°$ showing qualitatively similar CIs/zLLs sequences and phase boundaries as in Fig. 1d.

# Extended Data Figure 5

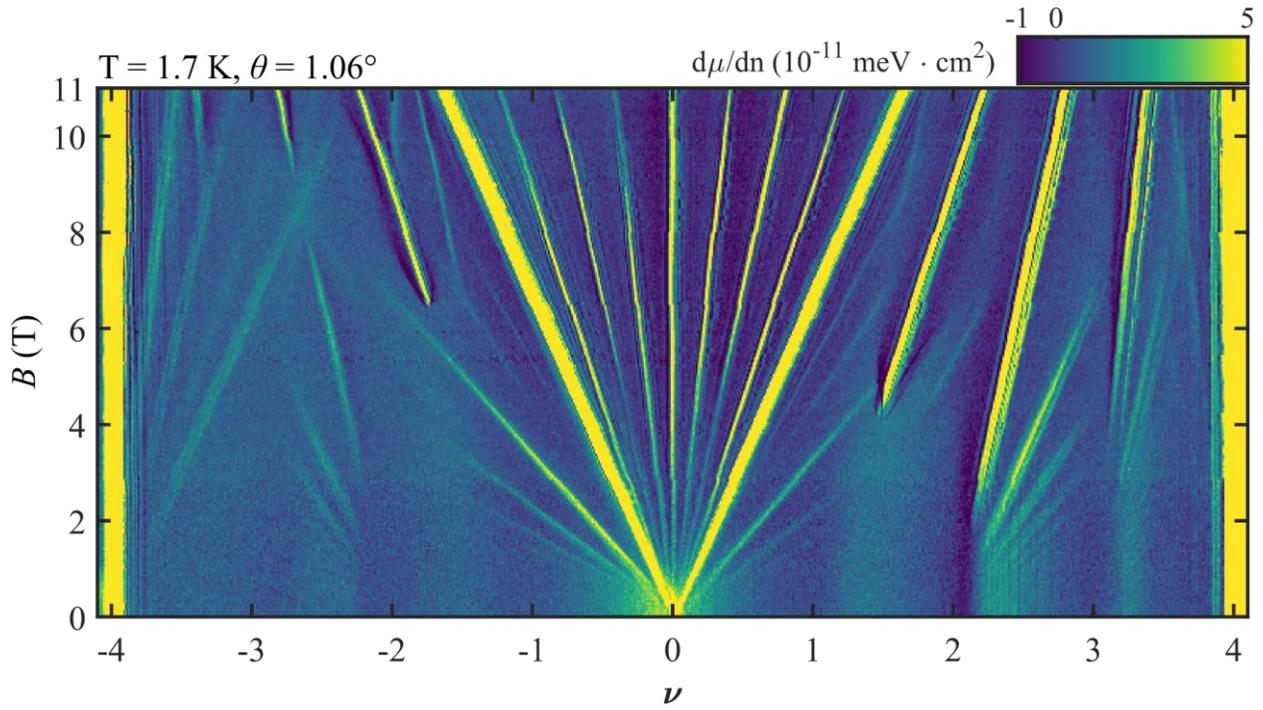

**Extended Data Fig. 5 | Landau fan at higher temperature.** $d\mu/dn$ measured at the location of Fig. 1d at temperature $T = 1.7$ K. The CIs/zLLs and negative $d\mu/dn$ features are thermally broadened and weakened, but the qualitative pattern is unchanged. Interestingly, negative $d\mu/dn$ features that mark phase transitions appear closer to the adjacent incompressible states.

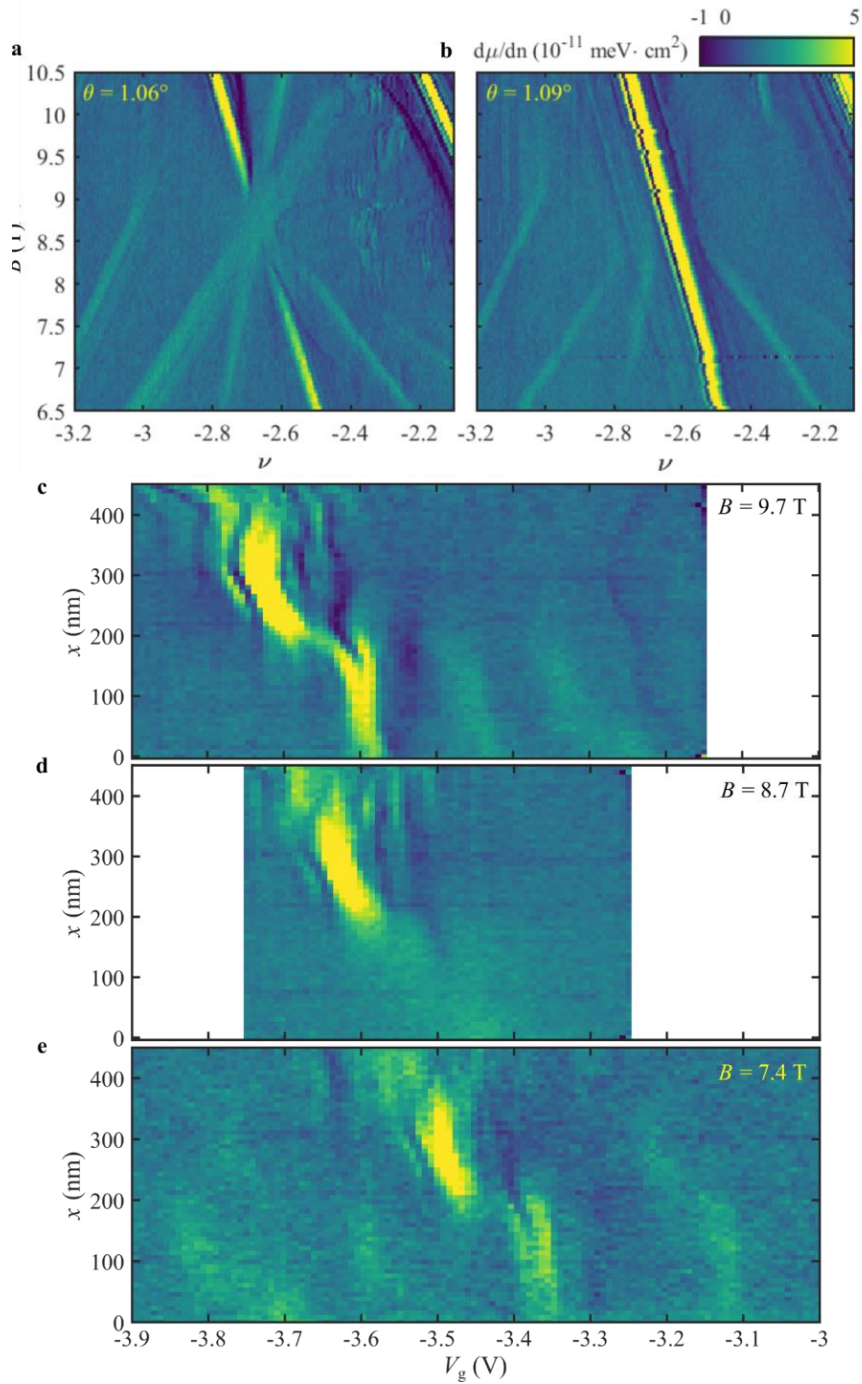

**Extended Data Fig. 6 | Spatial dependence and competing CIs. a-b,** Higher resolution Landau fans in the vicinity of ($\nu$ = -8/3, $B$ = 8.7 T) taken in the locations of Fig. 1d and Extended Data Fig. 4. The pattern of dominant CIs and detailed regions of negative $d\mu/dn$ are distinct in each location. **c-e,** Spatial linecuts (same trajectory as in Extended Data Fig. 1) taken at $B$ = 9.4, 8.7, and 7.4 T, respectively.

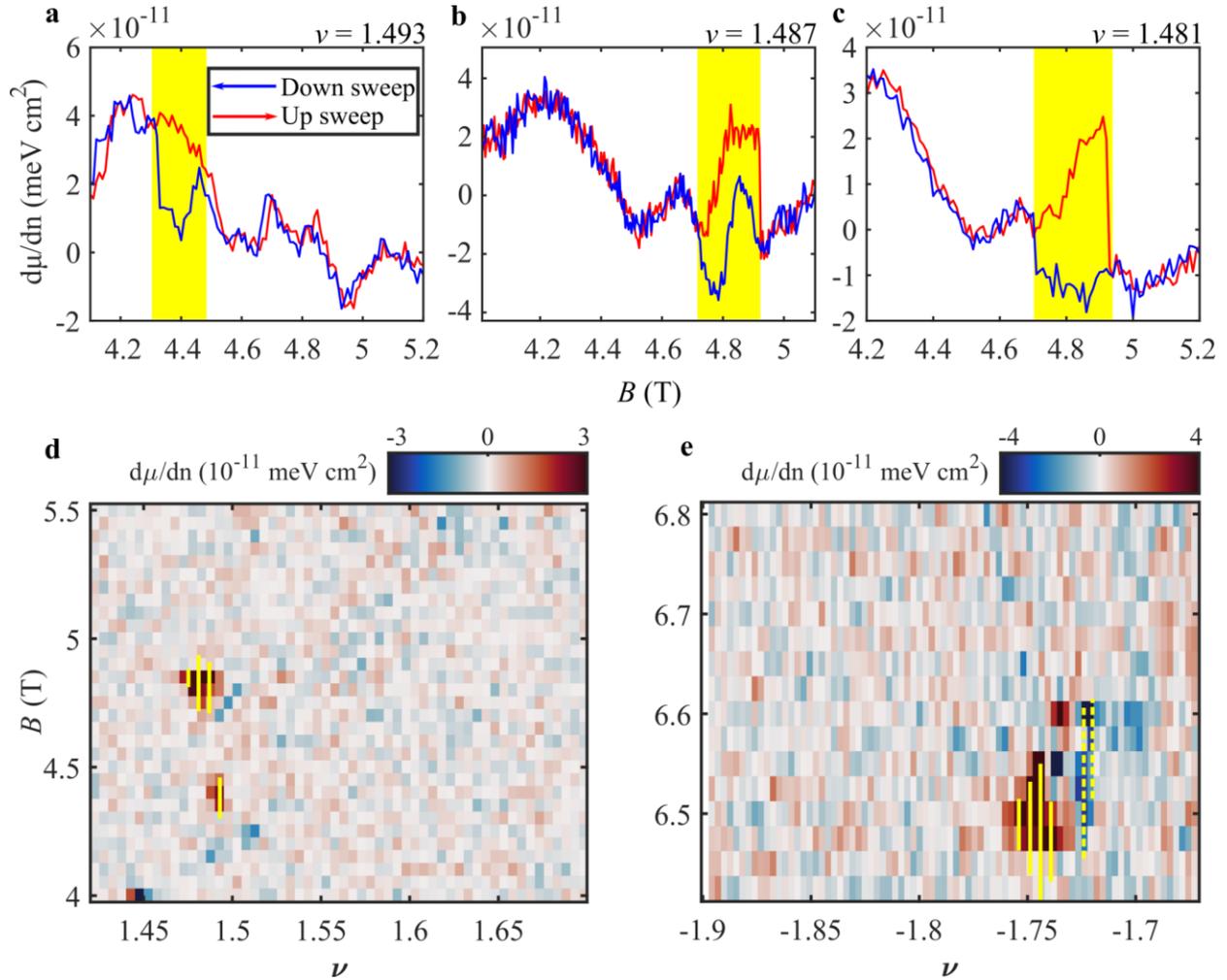

**Extended Data Fig. 7 | Hysteresis as a function of $B$. a-c,** $d\mu/dn$ measured as a function of both increasing (red) and decreasing (blue) $B$ at fixed densities. Hysteretic regions are highlighted in yellow. **d-e,** Hysteresis as a function of density, reproduced from Figs. 3a, d. Overlaid yellow lines indicate ranges over which we observe hysteresis as a function of $B$. Solid (dashed) lines indicate $d\mu/dn_{up} - d\mu/dn_{down} > 0 \,(< 0)$, respectively. The ranges of $\nu$ and $B$ in which hysteresis occurs are independent of which is swept as the fast axis.

**Extended Data Figure 8**

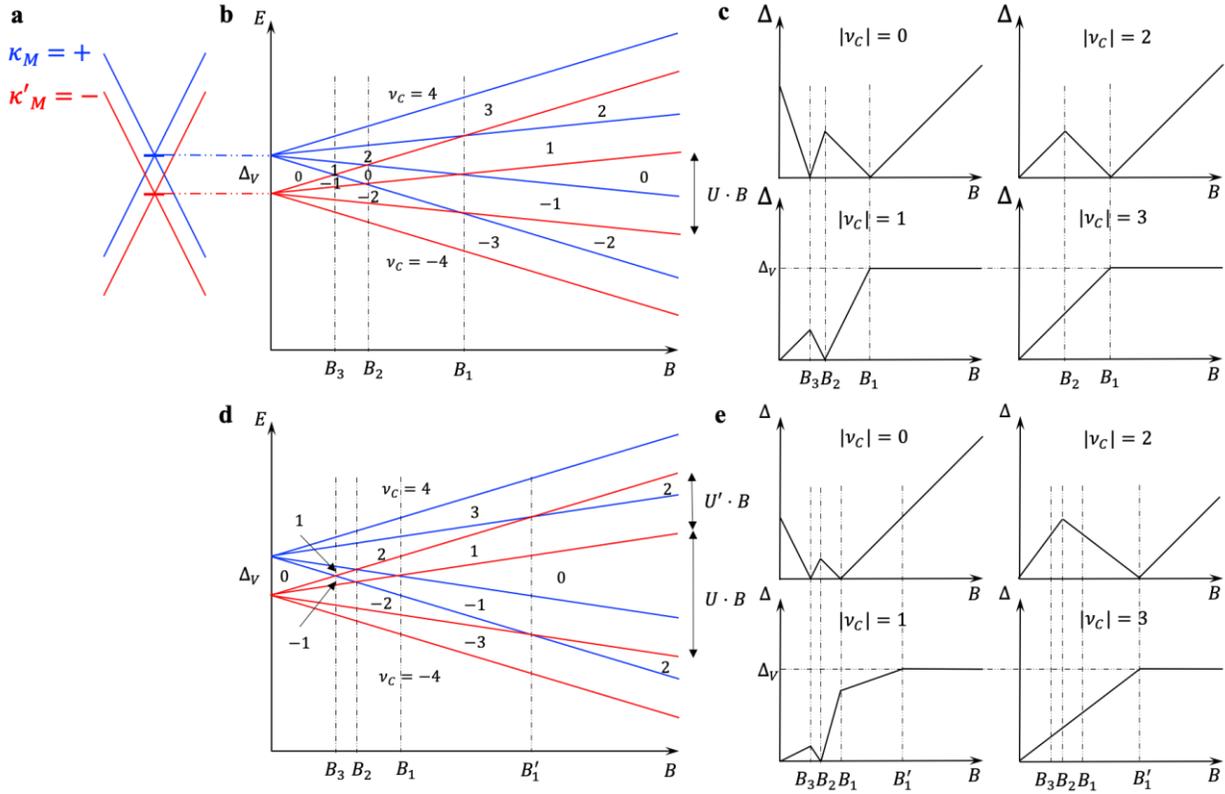

**Extended Data Fig. 8 | Phenomenological model of zLL evolution in the presence of mirror symmetry breaking. a,** Schematic of the moiré valley splitting $\Delta_v$ between $\kappa_M, \kappa'_M$ by $\Delta_v$. **b,** Evolution of the zLLs in the presence of inter-flavor Coulomb repulsion $U$ (ignoring single-particle effects). Four zLLs with different flavors are split linearly in $B$ within each moiré valley (red and blue, respectively), resulting in LL crossings. The corresponding Chern numbers are labeled within the gaps. **c,** $B$ dependence of zLL gaps derived from **b**. **d,** Evolution of the zLLs if the Zeeman coupling is different for spin and valley flavors. Such single-particle gaps lead to different effective interactions $U$ and $U'$ between LLs corresponding to different flavors within each moiré valley. This modifies the fields at which LL crossings occur, reducing $B_1, B_2$ and $B_3$ relative to the predicted field $B'_1$ at which the gaps at $|v_C| = 2$ close and those at odd integer fillings saturate. **e,** $B$ dependence of the zLL gaps derived from **d**.

**Extended Data Figure 9**

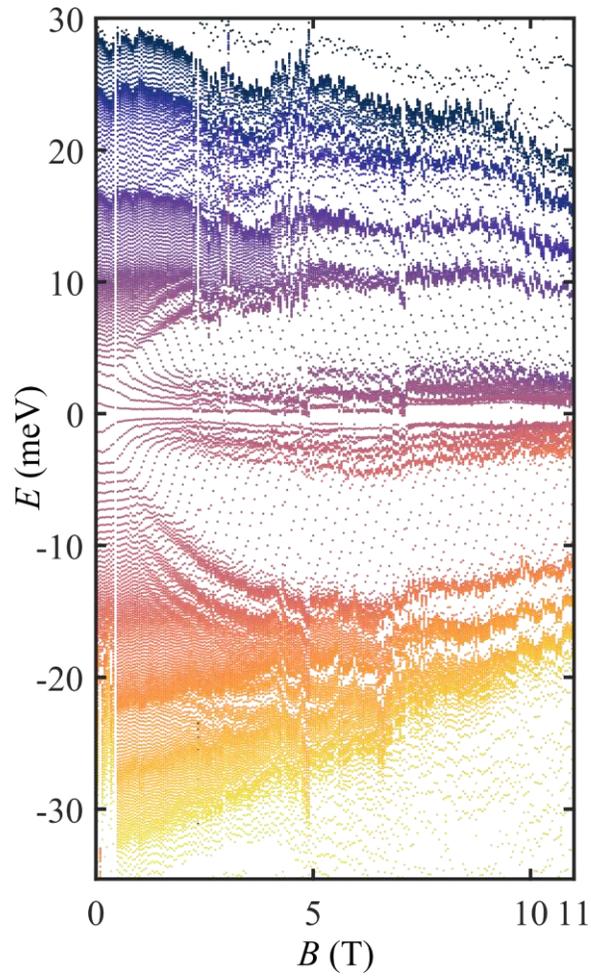

**Extended Data Fig. 9 | Hofstadter spectrum from a second location.** Correlated Hofstadter spectrum extracted from the data in Extended Data Fig. 4. The spectrum qualitatively agrees with that in Fig. 2c with a slightly smaller total bandwidth.

**Extended Data Figure 10**

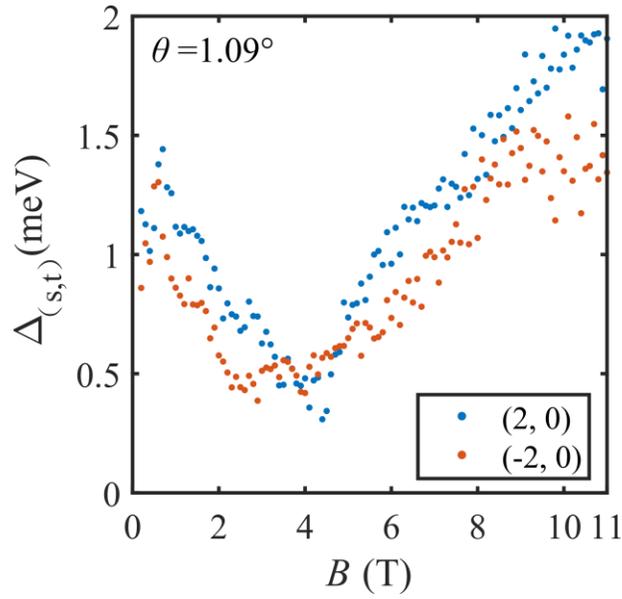

**Extended Data Fig. 10 | zLL gap evolution at a second location.** Thermodynamic gaps of $\nu_c = \pm 2$ zLLs extracted from the data presented in Extended Data Fig. 4.